# Solid-State $^{229}$Th Nuclear Laser with Two-Photon Pumping

Haowei Xu[1], Hao Tang[2], Guoqing Wang[1,3], Changhao Li[1,3], Boning Li[3,4],

Paola Cappellaro[1,3,4, †], and Ju Li[1,2, ‡]

[1] Department of Nuclear Science and Engineering, Massachusetts Institute of Technology, Cambridge, Massachusetts 02139, USA

[2] Department of Materials Science and Engineering, Massachusetts Institute of Technology, Cambridge, Massachusetts 02139, USA

[3] Research Laboratory of Electronics, Massachusetts Institute of Technology, Cambridge, MA 02139, USA

[4] Department of Physics, Massachusetts Institute of Technology, Cambridge, MA 02139, USA

* Corresponding authors: † pcappell@mit.edu,   ‡ liju@mit.edu

**Abstract**

The radiative excitation of the 8.3 eV isomeric state of thorium-229 is an outstanding challenge due to the lack of tunable far-ultraviolet (F-UV) sources. In this work, we propose an efficient two-photon pumping scheme for thorium-229 using the optonuclear quadrupolar effect, which only requires a 300 nm UV-B pumping laser. We further demonstrate that population inversion between the nuclear isomeric and ground states can be achieved at room temperature using a two-step pumping process. The nuclear laser, which has been pursued for decades, may be realized using a Watt-level UV-B pumping laser and ultrawide bandgap thorium compounds (e.g., ThF$_4$, Na$_2$ThF$_6$, or K$_2$ThF$_6$) as the gain medium.

**Introduction**. Thorium-229 ($^{229}$Th) nucleus exhibits a long-lived isomeric state ($^{229*}$Th) with an ultra-low energy of $\omega_{\text{is}} \approx 8.3$ eV above the ground state [1–3], in stark contrast to the typical nuclear excitation energies (keV to MeV) [4]. Such a low-energy isomeric state elicits considerable interest in understanding the underlying nuclear structure [5,6], the coupling between nuclear and electronic excitations [7–9], as well as in developing various applications such as creating a nuclear clock frequency standard [1,10–12] and determining fundamental physical constants [13,14].

$^{229}$Th also provides unique opportunities for building nuclear lasers, which were envisioned more than half a century ago, but have not been realized yet [15–17]. Besides the conceptual importance of nuclear lasing, nuclear lasers also feature short wavelengths and narrow linewidth that could facilitate various applications. $^{229}$Th nuclear lasers, if realized, could also offer a direct and convenient approach to optically pump $^{229}$Th nuclei, which is crucial for

many applications involving $^{229}$Th, including nuclear clock [1,10–12]. However, the construction of nuclear lasers is a demanding task that requires interdisciplinary research in nuclear physics, materials science, and photonics. A major challenge is that typical lasers require population inversion and hence efficient pumping to the excited states. However, it is notoriously difficult to pump to nuclear excited states. Historically, it has been proposed to use X-ray radiations, slow neutron capture, or other nuclear reactions to pump the nuclear excited states, all of which are not so efficient [15]. For $^{229}$Th, it is possible to populate the isomeric states optically, thanks to the small transition energy within the far-ultraviolet (F-UV) regime [17]. Unfortunately, F-UV sources resonant with the $^{229}$Th isomeric transition are still under development, and the direct laser excitation of $^{229}$Th has not been experimentally demonstrated yet [1]. Moreover, even if direct F-UV pumping becomes available, population inversion still cannot be achieved if only two states are involved, and at least one auxiliary state is necessary. For nuclear lasers, it is not straightforward to find the auxiliary state [17], because the energy gaps between nuclear spin sub-levels ($10^{-9} \sim 10^{-6}$ eV) are too small compared with typical laser frequency ($\sim 1$ eV), thus requiring prohibitively low temperatures [17]. Meanwhile, the energy gaps between nuclear orbital excited states are too large – the second excited state of $^{229}$Th is around 29 keV above the ground state – thus requiring demanding X-ray sources.

In this work, we propose a nuclear laser based on $^{229}$Th with a two-step two-photon pumping, which could potentially overcome the aforementioned challenges. As the gain medium, we suggest using Thorium compounds such as $Na_2ThF_6$, which can provide a high number density of $^{229}$Th and an ultrawide bandgap, $E_g > \omega_{is}$ [18]. The ultrawide bandgap also forbids internal conversion (IC) of the isomeric state [1,4], because the large ionization energy prevents this non-radiative nuclear decay process that excites and ejects a valence electron [19]. These properties are advantageous for the nuclear laser. Then, we introduce the two-photon pumping of the $^{229}$Th isomeric states based on the optonuclear quadrupolar (ONQ) effect [20,21], which is an efficient interface between two photons and the nuclei. The ONQ pumping requires only a near-ultraviolet laser operating at $\omega_{in} = \frac{\omega_{is}}{2} \approx 4.1$ eV, which is in the UV-B regime and could be much easier to build than a F-UV laser [22]. In contrast to the pumping scheme based on the electronic bridge (EB) effect [7,9,23–26], the ONQ pumping avoids the usage of lasers resonant with electronic transitions and can thus significantly suppress the heating in the solid-state gain medium. This is important when the $^{229}$Th density is high and the pumping laser is strong, whereby the heating power density will be high. Under a sub-Watt-level pumping laser, the ONQ pumping could be fast enough for the experimental observation of the radiative excitations of $^{229*}$Th, which has not been realized yet. We further propose a two-step pumping process to achieve population inversion at room temperature, taking advantage of the long relaxation time of $^{229}$Th nuclear states. We show that the peak power of the nuclear laser can reach Watt-level when the gain medium size is about $1\mu m \times 1\mu m \times 1mm$. By selecting different nuclear spin sub-levels, the nuclear laser can have tunable chirality as well [27]. The nuclear laser can have narrow linewidth, and can naturally match the resonance condition for optically pumping the $^{229}$Th nuclear clock. While $^{229}$Th can be pumped with other schemes under development or operation [2], the nuclear laser pumping may have its own advantage. For example, the central frequency of the nuclear laser can potentially be tuned with ultra-fine resolution using e.g., the Mössbauer effect. Further investigation is required to explore the potential applications of the nuclear laser.

**$^{229}$Th in ultrawide bandgap Th-compounds.** The radiative transitions between the isomeric (angular momentum $I_{is} = 3/2$) and ground states ($I_{gr} = 5/2$) of $^{229}$Th have both $M1$ (magnetic dipole) and $E2$ (electric quadrupole) channels. Some detailed information on the isomeric transition, including the selection rules, is summarized in Section 2 in Ref. [28] (Supplementary Materials, which also cites Refs. [5,6,29–38]). The spontaneous gamma-decay of $^{229*}$Th is dominated by the $M1$ process with a decay rate of $\gamma_{is}^{\gamma} \sim 10^{-4}$ Hz [4]. The IC, while fast [39], can be forbidden if the isomeric transition energy $\omega_{is}$ is below the electron ionization energy, so the nuclear transition does not have enough energy to kick out an electron. In this case, the total decay rate of $^{229*}$Th is $\gamma_{is}^{decay} = \gamma_{is}^{\gamma}$ [40]. It has been shown that using trapped ionized $^{229}$Th [41] or $^{229}$Th dopants in ultrawide bandgap compounds (e.g. CaF$_2$ [42]) can forbid IC. For nuclear lasers, it is desirable to have a large number density of $^{229}$Th. Hence, we instead suggest using natural Th-compounds, which can have a number density of up to $10^{26}$ m$^{-3}$ if $^{229}$Th is enriched to 1% isotopic abundance. Some candidate compounds are ThF$_4$, Na$_2$ThF$_6$, and K$_2$ThF$_6$, all of which have electronic bandgaps $E_g \gtrsim 10$ eV according to experiments [18] as well as our many-body $G_0W_0$ calculations (Section 1.1 in Ref. [28]).

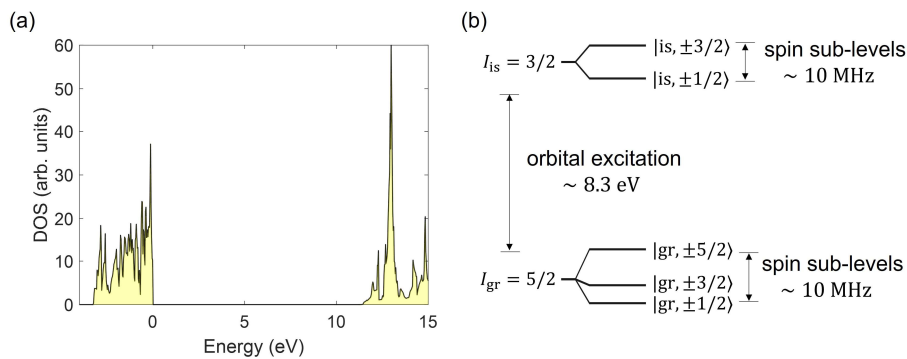

Figure 1. (a) Electronic density of states of Na$_2$ThF$_6$ from $G_0W_0$ calculations. (b) Nuclear energy level diagram of $^{229}$Th in Na$_2$ThF$_6$.

We will use Na$_2$ThF$_6$ as an example. According to our calculations, the electronic bandgap of Na$_2$ThF$_6$ is $E_g \approx 11$ eV (Figure 1a), yielding $\omega_{is} < E_g < \frac{3}{2}\omega_{is}$. The splitting between nuclear spin sub-levels due to the nuclear quadrupolar interaction is on the order of 10 MHz (40 neV, Figure 1b), equivalent to a temperature of mK. Hence, under ambient conditions, the five ground-state sub-levels are almost equally populated. In contrast, $\omega_{is}$ is much greater than the thermal energy, and the isomeric sub-levels should have zero population at thermal equilibrium. Another important parameter for the nuclear laser is the drift (inhomogeneous broadening) $\gamma_{is}^{drift}$ of the isomeric transition energy in solid-state compounds, which could result from magnetic interactions, temperature, and strain effects [31]. The magnetic dipole interaction between nearby nuclei is on the order of kHz [43], while our calculations (Section 1.1 in Ref. [28]) indicate that $\gamma_{is}^{drift}$ can be kept below 10 kHz if the variance of strain (temperature) is below $10^{-2}$ % (1 K). Hence, we will assume $\gamma_{is}^{drift} \sim 10$ kHz in the following. The small drift in the isomeric transition energy also indicates that the nuclear laser can have a narrow linewidth.

**Two-photon pumping via the optonuclear quadrupolar effect.** As discussed before, the pumping of the nuclear isomeric excited state is a key challenge for nuclear lasers. In this section, we demonstrate the two-photon pumping of $^{229}$Th based on the ONQ effect [20,21]. Specifically, the nuclear state can be influenced by the nuclear quadrupolar ($E2$) interaction $\mathcal{H}_{E2} = \mathcal{M}_{E2}\mathcal{V}$, which is an electromagnetic interaction between the nuclear electric quadrupolar moment $\mathcal{M}_{E2}$ and the electric field gradient (EFG) $\mathcal{V}$ at the site of the nucleus. External fields can modulate $\mathcal{V}$, which can in turn control the nuclear states. In fact, the gamma-decay through the $E2$ channel is the consequence of the oscillating EFG of a resonant photon [44]. However, the EFG of a VUV photon is too weak, and thus the $E2$ channel for gamma-decay is inefficient compared with the $M1$ channel for bare $^{229*}$Th in vacuum [6].

The situation is different when electrons come into play. The electric field generated by electrons can vary by $\Delta\mathcal{E}_e \gtrsim 1$ V/Å over the atomic scale $a_0$, with $a_0$ the Bohr radius. Hence, the EFG $\mathcal{V}_e$ generated by electrons can reach $\mathcal{V}_e \sim \frac{\Delta\mathcal{E}_e}{a_0} \sim 1$ V/Å$^2$, leading to a strong nuclear quadrupolar interaction (MHz to GHz). When the electronic states are perturbed, the change in the nuclear quadrupolar interaction is proportionally strong. This fact helps explain the fast IC of $^{229*}$Th through the $E2$ channel [19].

Particularly, the electronic states can be perturbed by two-photon transitions. This is the origin of various well-known second-order nonlinear optical effects – two photons drive the electronic orbital motions, which in turn generate e.g., electromagnetic waves (sum or difference frequency generation) or phonons (Raman scattering). Similarly, electronic orbital motions can also generate oscillating EFG and hence oscillating nuclear quadrupolar interaction, which can influence the nuclear states. This is the ONQ effect [20,21], which can be described by the Hamiltonian $\mathcal{H}_{E2}^{ONQ}(t) = \sum_{ij} \mathcal{D}_{ij,\pm}^{pq} \mathcal{E}_p \mathcal{E}_q e^{i(\omega_p \pm \omega_q)t} + h.c.$, where $\mathcal{E}_{p,q}$ and $\omega_{p,q}$ are the electric field strength and the frequency of the two photons $p$ and $q$, respectively. The $\pm$ sign indicates the sum (+) or difference (−) frequency process. The response function $\mathcal{D}_{ij,\pm}^{pq} = \mathcal{M}_{E2} \frac{\partial^2 \mathcal{V}}{\partial \mathcal{E}_p \mathcal{E}_q}$ can be expressed as [20,21]

$$\mathcal{D}_{ij,\pm}^{pq} = \mathcal{M}_{E2} \sum_{mnl} \frac{[\mathcal{V}_{ij}]_{mn}}{\omega_{mn} - (\omega_p \pm \omega_q)} \times \left\{ \frac{f_{lm}[r_p]_{nl}[r_q]_{lm}}{\omega_{ml} - \omega_p} - \frac{f_{nl}[r_q]_{nl}[r_p]_{lm}}{\omega_{ln} - \omega_p} \right\} + (p \leftrightarrow q) \quad (1)$$

where $(p \leftrightarrow q)$ indicates the exchange of the $p$ and $q$ subscripts. $[r_i]_{nl} \equiv \langle n|r_i|l \rangle$ and $[\mathcal{V}_{ij}]_{mn} = \frac{e}{4\pi\varepsilon_0} \langle m | \frac{3r_i r_j - \delta_{ij} r^2}{r^5} | n \rangle$ are the electron position and EFG operators, respectively, where $m, n, l$ label the electronic states, and $\varepsilon_0$ is the vacuum permittivity. $\omega_{mn}$ ($f_{mn}$) is the energy (occupation) differences between two electronic states $|m\rangle$ and $|n\rangle$ (Planck constant $\hbar = 1$). Here, we focus on the sum-frequency term $e^{i(\omega_p + \omega_q)t}$, which can pump the electrons to a (virtual) excited state with an energy of $\omega_p + \omega_q$. Then, the nuclear excitation can be realized by a swap process between the electronic and nuclear excitations – electrons (virtually) jump back to the ground state, while the nucleus jumps to the isomeric state (Figure 2a). This process is enabled by the electron-nuclear interactions, which can have $M1$, $E2$, and other higher-order channels. In the case of Na$_2$ThF$_4$, there are no net electron spins, so the $M1$ channel is absent. To leading order, we only need to consider the $E2$ channel.

When $\omega_p + \omega_q < E_g$, the electronic transition is virtual, but the nuclear transition can be a real resonant transition when $\omega_p + \omega_q = \omega_{is}$. When the laser frequencies ($\omega_p$, $\omega_q$ or $\omega_p + \omega_q$) are resonant with an electronic transition, the electrons can be resonantly pumped to electronic excited states, and the $\mathcal{D}$ tensor will be substantially enhanced. In this case, the ONQ effect is in principle equivalent to the EB process [7,9,23–26]. We would like to emphasize that a unique advantage of the ONQ effect is that the laser frequencies can be off-resonant with electronic transitions (below bandgap), which can significantly suppress the one-photon absorption of laser energy and the resultant heating. This difference with the EB process is particularly important when the number density of Th is high and the pumping laser is strong, both of which are desirable for the solid-state nuclear laser. Additionally, the EB process is not favorable in an ultra-wide bandgap thorium compound, as the one-photon resonant transition requires a laser with frequency >10 eV, which is hard to construct. Therefore, we believe the ONQ effect can be more advantageous than the EB process regarding building the nuclear laser.

For an order-of-magnitude estimation, we only consider the $(m, n, l)$ pair that has $\omega_{mn}, \omega_{ml}$ close to $E_g$, which makes a major contribution to $\mathcal{D}$. We also use $\left\langle m \left| \frac{3r_i r_j - \delta_{ij} r^2}{r^5} \right| n \right\rangle \approx \frac{1}{a_0^3}$ and $[r_i]_{mn} \approx a_0$, as the spatial distribution of the electronic states is characterized by $a_0$. We also set $\omega_p = \omega_q = \omega_{\text{in}} = \frac{\omega_{\text{is}}}{2}$. Finally, one has $\mathcal{D}_+ \sim \mathcal{M}_{E2} \frac{g_S e^3}{4\pi\varepsilon_0 a_0} \frac{1}{(E_g - \omega_{\text{is}})(E_g - \omega_{\text{is}}/2)}$, where $g_S = 2$ is the electron spin degeneracy.

The pumping rate to the isomeric state is $R = \frac{4|\langle \text{gr}, m_{\text{gr}} | \mathcal{D}_+ | \text{is}, m_{\text{is}} \rangle|^2 \mathcal{E}^4}{\Gamma_{\text{pump}}}$ with $\Gamma_{\text{pump}} \sim \gamma_{\text{is}}^{\text{decay}} + \gamma_{\text{is}}^{\text{drift}} + \kappa_{\text{in}}$, where $\kappa_{\text{in}}$ is the linewidth of the pumping laser. One has $R[\text{Hz}] \sim 10^{-5} \times \mathcal{E}^4[\text{MV}^4 \cdot \text{m}^{-4}]$ when $\Gamma_{\text{pump}}$ is on 10 kHz scale. When $\mathcal{E} = 1\,\text{MV} \cdot \text{m}^{-1}$, one has $R \sim 10^{-5}$ Hz. If a $[10\mu\text{m}]^3$ Na$_2$ThF$_6$ sample with a $^{229}$Th number density of $10^{26}\,\text{m}^{-3}$ is used, then there will be $\sim 10^6$ excitations to, and $\sim 10^2$ radiative decays from the isomeric state per second. This could be fast enough for the experimental observation of the nuclear radiative emission.

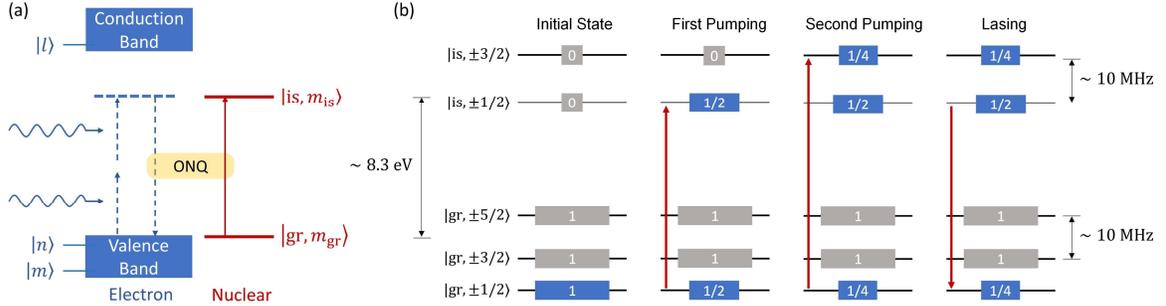

**Figure 2** (a) Two-photon pumping scheme based on the ONQ effect. Two-photons pumps a (virtual) electronic excitation, which is then swapped to a real nuclear excitation through the nuclear quadrupolar interaction. (b) The two-step pumping scheme to achieve population inversion. Numbers in the boxes indicate normalized populations. States with grey boxes do not participate in the nuclear pumping/lasing process.

**Population inversion under two-step pumping.** While the two-photon process discussed above can pump the isomeric state, it cannot lead to a population inversion, because it only involves two nuclear states. For population inversion, at least one other auxiliary state is necessary [33]. The second nuclear orbital excited state is not an ideal choice because its high energy (29 keV) necessitates demanding X-ray sources. A more practical option is to use the nuclear spin sub-levels. However, the energy splitting between these spin sub-levels is too small. Hence, if the common one-step pumping scheme is used, then an effective population can be achieved only under a cryogenic temperature of millikelvin (Section 3 in Ref. [28]).

For room-temperature nuclear lasing, we propose a two-step pumping process (Figure 2b). Initially, the system is at thermal equilibrium, so the ground state sub-levels have (approximately) the same population (normalized to $f = 1$), while the isomeric states are empty ($f = 0$). For the first step of the pumping process, we use a two-photon pumping laser resonant with the $|gr, \pm 1/2\rangle \leftrightarrow |is, \pm 1/2\rangle$ transition. Provided the pumping rate is large enough ($R \gg \gamma_{is}^{decay}$), the $|is, \pm 1/2\rangle$ state will have almost the same population ($f = \frac{1}{2}$) as the $|gr, \pm 1/2\rangle$ state when the pumping saturates. Then, we switch to a second pumping laser resonant with the $|gr, \pm 1/2\rangle \leftrightarrow |is, \pm 3/2\rangle$ transition, which again equalizes the final population ($f = \frac{1}{4}$) on these two states. This results in the population inversion between $|is, \pm 1/2\rangle$ ($f = \frac{1}{2}$) and $|gr, \pm 1/2\rangle$ ($f = \frac{1}{4}$), and nuclear lasing between these two states would start when the laser resonator is tuned on resonance. A greater population inversion can be achieved if a multi-step pumping sequence is used (Section 3.2 in Ref. [28]). A caveat is that the two-photon pumping rate $R$ should be faster than the nuclear spin relaxation, which tends to equalize the population among nuclear spin sub-levels at room temperature and destroy the population inversion. Considering that the nuclear spin relaxation rate is usually on the order of Hz at room temperature [45,46], a pumping rate of $R = 100$ Hz should suffice. The corresponding pumping electric field is $\mathcal{E}_{in} \approx 0.56$ MV/cm, and the laser power is $P_{in} \approx 4.2$ W when the spot size is $[1\mu m]^2$. With a pumping rate of $R = 100$ Hz, it takes $0.01 \sim 0.1$ s to reach the two-level saturation. In addition, we remark that the linewidth of the pumping laser should be smaller than the splitting between nuclear spin sub-levels (around 10 MHz in Na$_2$ThF$_6$), so that it can be resonant with just one nuclear transition at a time.

**Experimental setup and performance of the nuclear laser.** Next, we discuss the basic experimental setup of the nuclear laser. We assume the gain medium (e.g., Na$_2$ThF$_6$) has a volume $V = Sl$, with $S$ the cross-sectional area and $l$ the length. The total number of *active* $^{229}$Th nuclei is $N_{Th} = f_n \rho_n V$, where $\rho_n$ is the number density of $^{229}$Th, while $f_n \approx 1$ is the Lamb–Mössbauer factor (also known as the Debye-Waller factor) [47]. For clarity, we fix $f_n \rho_n \sim 10^{26}$ m$^{-3}$, which can be achieved when $^{229}$Th is enriched to $\sim 1\%$ abundance. Note that an abundance of $^{229}$Th exceeding 75% has been realized before [48]. Given a gain medium with a $1\mu m \times 1\mu m \times 1mm$ dimension (see below), the total weight of $^{229}$Th nuclei is around $4 \times 10^{-11}$ gram, much smaller than the current global stock of $^{229}$Th (tens of grams [1]). The total pumping rate is $\mathcal{R} \equiv N_{Th}R \propto Sl\mathcal{E}^4$, while the total power of the pumping laser is $P_{in} \propto S\mathcal{E}^2$, yielding $\mathcal{R} \propto \frac{P_{in}^2 l}{S}$. Hence, a smaller $S$ improves $\mathcal{R}$, as it is typical for nonlinear two-photon processes [34]. Considering that the wavelength of the pumping laser is around 300 nm, we suggest using $S = [1\mu m]^2$. On the other hand, the length $l$ should not be too small, because the performance of the pulsed nuclear laser, including peak power and number of photons per pulse, increases with $l$. For demonstrative purposes, we will use $l = 1$ mm hereafter, but we have not optimized these parameters. The gain medium is thus a nanowire of size $1\mu m \times 1\mu m \times 1mm$. Actually, nanowire lasers have been demonstrated before based on electronic transitions [49].

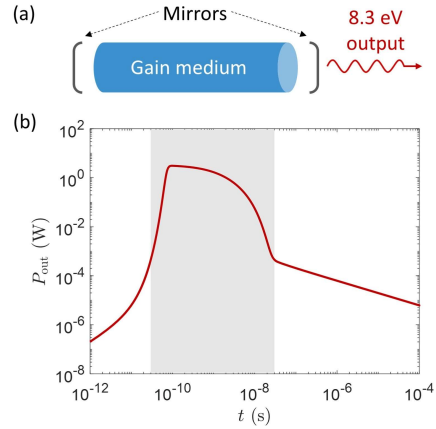

**Figure 3** (a) A simplified setup of the nuclear laser. (b) Time evolution of the output power of the nuclear laser. The parameters of the nuclear laser are described in the main text. The shaded area indicates a nuclear laser pulse.

We further assume that the nanowire gain medium is confined by two mirrors on two sides, which forms an optical cavity (Figure 3a). The left mirror is a total reflector with 100% reflectivity for the 8.3 eV cavity photon, while the right mirror is a partial reflector with a transmissivity of $T$, which serves as the output channel of the cavity photons. The stimulated emission (absorption) rate of the cavity photons can be expressed as $K = \frac{4|\langle gr, m_{gr}|\mathcal{M}_{M1}|is, m_{is}\rangle|^2 \mathcal{B}_{zpf}^2}{\Gamma_0}$, where $\Gamma_0 \sim \gamma_{is}^{decay} + \gamma_{is}^{drift}$ is the total broadening of the isomeric states, $\mathcal{M}_{M1}$ is the nuclear magnetic dipole transition dipole, while $\mathcal{B}_{zpf} = \sqrt{\frac{\mu_0 \omega_{is}}{2V}}$ is the zero-point magnetic field of the cavity photon with $\mu_0$ the vacuum permittivity. Note that we only consider the $M1$ channel, as it is much more efficient than the $E2$ channel for radiative transitions that do not involve electrons [32].

The performance of the nuclear laser can be evaluated using the semi-classical rate equations (details in Section 4.1 in Ref. [28]). A typical time evolution of the output power $P_\text{out}$ is plotted in Figure 3b, where one can clearly see a nuclear laser pulse with a peak power of above 1 Watt and a duration of about 10 ns (shaded area in Figure 3b). A total of $4.6 \times 10^9$ nuclear gamma photons (8.3 eV) will be emitted per pulse. Because these highly coherent and collinear photons come from the stimulated emission of the nuclear excited states, our device would qualify as a gamma-ray laser ("graser"). In Section 4.2 of Ref. [28], we also show that the losses and temperature rise in the gain medium are minor and would not influence the operation of the nuclear laser.

Here we would like to remark on some potential challenges in constructing the nuclear laser proposed in this work. First, the nuclear lasing requires fast pumping of the nuclear isomeric state, so a ~ 4.1 eV UV-B laser with narrow linewidth and high power, which are assumed to be 10 kHz and 1 W in the discussions above, would be necessary. Such a laser would be challenging to build. But we expect it could be easier than building an ~ 8.3 eV F-UV laser. Additionally, our proposal implicitly assumes that the isomeric transition energy $\omega_\text{is}$ is known with high precision, which has not been realized yet. Fortunately, $\omega_\text{is}$ has been measured with increasing precision [50] recently. Potentially, the two-photon ONQ pumping proposed in the work can be used to measure $\omega_\text{is}$. To this purpose, one needs a pumping laser with tunable frequency. On the other hand, the pumping rate does not necessarily need to be high, so a laser with relatively wide linewidth and low output power may be sufficient.

In summary, we propose a two-photon pumping scheme to populate the long-lived nuclear isomers $^{229*}$Th based on the ONQ effect in solid crystals. This pumping scheme could be used in nuclear clocks based on $^{229}$Th as well. We further propose a nuclear gamma-ray laser (as this emission originates from nuclear isomeric transition) that utilizes ultrawide bandgap $^{229}$Th-compounds as the gain medium and a two-step, two-photon scheme to achieve population inversion at room temperature. Pulsed nuclear lasing should be realizable with a Watt-level pumping laser. The nuclear laser with narrow linewidth might be useful for various applications in e.g., nano-imaging, nuclear clock, and quantum information processing.

## Acknowledgment

We acknowledge support by DTRA (Award No. HDTRA1-20-2-0002) Interaction of Ionizing Radiation with Matter (IIRM) University Research Alliance (URA) and Office of Naval Research MURI through Grant No. N00014-17-1-2661. H. X. thanks Meihui Liu for helping with figure production.